# How big was Galileo's impact? Percussion in the Sixth Day of the "Two New Sciences"

*To Giovanna Geymonat, in memoriam*

Antonio Giannella-Neto and Frederico C. Jandre
agn@peb.ufrj.br  jandre@peb.ufrj.br

Biomedical Engineering Programme
COPPE - Federal University of Rio de Janeiro
Rio de Janeiro - Brazil

**Abstract**

The Giornata Sesta about the Force of Percussion is a relatively less known Chapter from the Galileo's master's piece "Discourse about Two New Sciences". It was first published lately (1718), long after the first edition of the Two New Sciences (1638) and Galileo's death (1642).

The Giornata Sesta focuses on how to quantify the percussion force caused by a body in movement, and describes a very interesting experiment known as "the two-bucket experiment".

In this paper, we review this experiment reported by Galileo, develop a steady-state theoretical model, and solve its transient form numerically; additionally, we report the results from one real simplified analogous experiment. Finally, we discuss the conclusions drawn by Galileo -- correct, despite a probably unnoticeable imbalance --, showing that he did not report the thrust force component in his setup -- which would be fundamental for the correct calculation of the percussion force.

## 1. Introduction

First, the experiment will be reported in its original form [1] as translated by Stillman Drake [2]. To help to understand the setup, Figure 1 shows a photograph of a model built with dimensions similar to those described by Galileo [3].

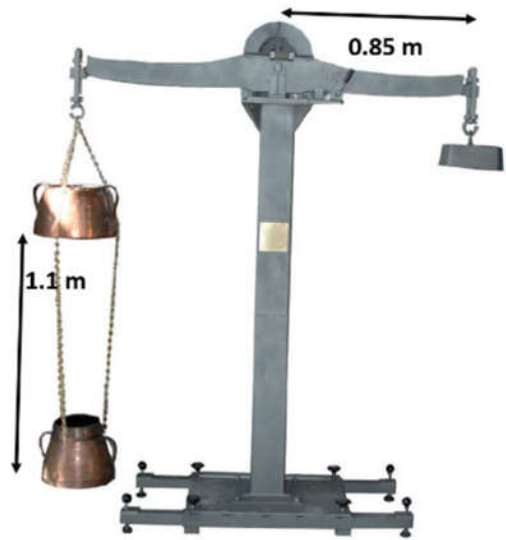

Figure 1 – A model of the two-bucket setup: the upper bucket is filled with water and has a hole on its bottom. Initially, the hole is closed, and the balance scale is in equilibrium. When the hole is opened, the water flows into the lower bucket. Initially, the scale goes out of balance, moving clockwise, and after the jet of water touches the lower bucket, the equilibrium is restored. This model was found at the University of Pisa, Italy [3].

Report of the experiment in the words of Aproino, a Galileo's disciple:

> "He took a very sturdy rod, about three braccia long, pivoted like the beam of a balance, and he suspended at the ends of these balance-arms two equal weights, very heavy One of these consisted of copper containers, that is, of two buckets, one of which hung at the said extremity of the beam and was filled with water. From the handles of this bucket hung two cords, about two braccia each in length, to which was attached by its handles another like bucket, but empty, this hung plumb beneath the bucket already described as filled with water. At the end of the other balance-arm he hung a counterweight of stone or some other heavy material, which exactly balanced the weight of the whole assembly of buckets, water, and ropes. The bottom of the upper bucket had been pierced by a hole the size of an egg or a little smaller, which hole could be opened and closed. Our first conjecture was that when the balance rested in equilibrium, the whole apparatus having been prepared as described, and then the [hole in the] upper bucket was un-stoppered and the water allowed to flow, this would go swiftly down to strike in the lower bucket, and we conceived that the adjoining of this impact must add to the [static] moment on that side, so that in order to restore equilibrium it would be necessary to add more weight to that of the counterpoise on the other arm. This addition would evidently restore and offset the new force of impact of the water, so that we could say that its momentum was equivalent to the weight of the ten or twelve pounds that it would have been necessary [as we imagined] to add to the counterweight.
> SAGREDO. This scheme seems to me really ingenious, and I am eagerly waiting to hear how the experiment succeeded.
> APROINO. The outcome was no less wonderful than it was unexpected by us. For the hole being suddenly opened, and the water commencing to run out, the balance did indeed tilt toward the side with the counterweight, but the water had hardly begun to strike against the bottom of the lower bucket when the counterweight ceased to descend, and commenced to rise with very tranquil motion, restoring itself to equilibrium while water was still flowing, and upon reaching equilibrium it balanced and came to rest without passing a hair breadth beyond.
> SAGREDO. This result certainly comes as a surprise to me. The outcome differed from what I had expected, and from which I hoped to learn the amount of the force of impact. Nevertheless it seems to me that we can obtain most of the desired information. Let us say that the force and moment of impact is equivalent to the moment and weight of whatever amount of falling water is found to be suspended in the air between the upper and lower buckets, which quantity of water does not weigh at all against either upper or lower bucket. Not against the upper, for the parts of water are not attached together, so they cannot exert force and draw down on those above, as would some viscous liquid, such as pitch or lime, for example. Nor [does it weigh] against the lower [bucket], because the falling water goes with continually accelerated motion, so its upper parts cannot weigh down on or press against its lower ones. Hence it follows that all the water contained in the jet is as if it were not in the balance. Indeed, that is more than evident, for if that [intermediate] water exerted any weight against the buckets that weight together with the impact would greatly incline the buckets downward, raising the counterweight, and this is seen not to happen. This is again exactly confirmed if we imagine all the water suddenly to freeze, for then the jet, made into solid ice, would weigh with all the rest of the structure, while cessation of the motion would remove all impact.

APROINO. Your reasoning conforms exactly with ours—immediately after the experiment we had witnessed. To us also, it seemed possible to conclude that the speed alone, acquired by the fall of that amount of water from a height of two braccia, without [taking into account] the weight of this water, operated to press down exactly as much as did the weight of the water, without [taking into account] the impetus of the impact. *Hence if one could measure and weigh the quantity of water hanging in air between the containers, one might safely assert that the impact has the same power to act by pressing down as would be that of a weight equal to the ten or twelve pounds of falling water (italic by the authors).*

SALVIATI. This clever contrivance much pleases me, and it appears to me that without straying from that path, in which some ambiguity is introduced by the difficulty of measuring the amount of this falling water, we might by a not unlike experiment smooth the road to the complete understanding which we desire."[2]

In short, after the bucket's hole has been opened the scale goes out of balance, moving clockwise; as the water column touches the bottom of the lower bucket, the equilibrium is restored. Galileo interpreted this result as if the weight of the water travelling between the containers was identical to the force of percussion.

## 2. Demonstration for a steady-state condition

Let us consider the simplest case where the upper bucket is permanently full of water and, therefore, after the hole is opened, the stream of falling water reaches a steady-state condition. Scalar equations are employed in this demonstration, considering only the effects on the vertical axis.

Consider the vertical distance $H$ between the bottom of the upper bucket and the surface of the water at the lower bucket, the velocity $v_0$ of the water stream at the upper bucket hole, the specific mass $\rho$, the acceleration of gravity $g$ and the area $A_0$ of the orifice. Neglecting frictional forces and the presence of a vena contracta profile close to the hole, and applying Torricelli's theorem plus the continuity equation, the volume of water travelling between the containers is:

$$Volume_{water} = \int_0^H \frac{v_0 A_0}{\sqrt{v_0^2 + 2gh}} dh = \frac{v_0 A_0}{g}\left[\sqrt{v_0^2 + 2gH} - |v_0|\right] \qquad (1)$$

The corresponding weight of this volume is

$$Weight_{water} = \rho \cdot g \cdot Volume_{water} = \rho v_0 A_0 \left[\sqrt{v_0^2 + 2gH} - |v_0|\right] \quad (2)$$

The percussion of the water as it hits the surface of the lower bucket at a speed $v_{final}$, taking into account the total change in linear momentum, is

$$Percussion_{water} = mass\_flowrate \cdot v_{final} = \rho v_0 A_0 \sqrt{v_0^2 + 2gH} \quad (3)$$

Indeed, Equations 2 and 3 differ, and the percussion, in any circumstance, is larger than the weight of water in free fall. The difference between them is the term $\rho v_0^2 A_0$ that matches exactly the thrust of the water jet, experienced by the upper bucket. However, the sum of the forces applied to the system of two buckets after the water jet hits the lower bucket is the same as in the starting static condition. Thus, in this steady-state case, the equilibrium reported by Galileo is correct, despite a non-negligible error (see results) in the magnitude of the percussion force.

## 3. Materials and Methods

### 3.1 The real Galilean experiment

In the real case, the upper bucket is filled with a finite amount of water, thus, upon opening the hole, the water level decreases continuously as the bucket drains. A dynamic steady-state condition is never reached while the water is falling and, at any given instant, the water jet presents a continuum of flow rates that decrease over time and along its height.

The magnitude of the forces (percussion, thrust and water jet weight) in the real case were calculated by numerical simulation, with a discrete time interval of 0.1 ms, by a computer program written in Matlab (MathWorks). The program was previously tested in the above mentioned steady-state case, for validation.

The following parameters were considered during simulations: cylindrical upper and lower buckets, both with diameters db=0.30 m; initial water height at the upper bucket hbs=0.3 m; a circular hole with diameter hd=0.015 m; vertical distance between the bottoms of the buckets H=1.10 m. All results were calculated in SI units, with $\rho$=1000 kg.m$^{-3}$ and $g$=9.8 m.s$^{-2}$. These geometrical dimensions were similar to those described by Galileo, although the size of the orifice, as large as one egg in his words, is rather uncertain.

### 3.2 A physical test

A simplified version of the physical experiment was performed. Instead of the setup reported by Galileo and shown in Figure 1, in which the static equilibrium may be hindered by the swing of the balance arm, two buckets with capacities of 4 L each were firmly attached to one another with two metal rods. This assembly was laid on a weighing scale with a resolution of 0.001 kg (SilverCrest IAN104357, Germany), corresponding to a force of approximately 0.01 N. The measured dimensions were dB= 0.17 m; hbs=0.12 m; hd=0.0019 m; H=0.70 m. At the beginning of the experiment, the scale was tared; a video of the weighting scale digital display was recorded until the upper bucket was empty.

## 4. Results

### 4.1 Numerical Simulations

As expected, the simulation considering steady-state conditions shows that the percussion force equals the weight of the travelling water plus the thrust force in the upper bucket. These latter two terms represent, shortly after the opening of the hole, respectively 53% and 46% of the percussion. As a consequence, taking percussion to be equal to the weight of water in the air means a large underestimation of the magnitude of the former.

### 4.2 The real Galilean experiment

The simulation of the real case, in which the upper bucket drains completely over time, evidently shows that percussion, as well as the weight of water in the air and the thrust, continuously decreased during the experiment. However, it is interesting to note that the balance found in the steady-state experiment is almost achieved in this case. Figure 2a graphs the percussion, the thrust and the weight of water in the air throughout the experiment, and Figure 2b shows percussion plus thrust, minus the weight of water in the air, which peaks at about 0.12 N at the beginning of the test and corresponds to a weight of roughly 12 grams pulling the balance toward the buckets.

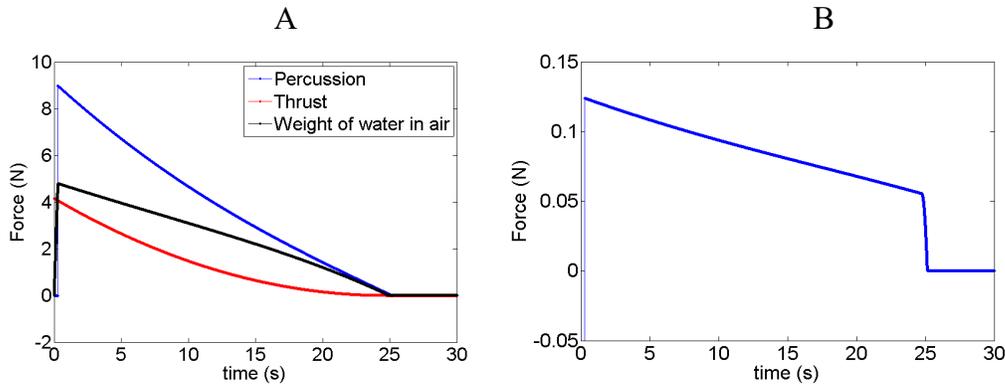

Figure 2: Magnitude of the forces during the simulated Galilean two-bucket experiment; A: Percussion, thrust and weight of water in the air; B: Net resulting upward force applied to the arm of the balance. The net force is slightly less than during rest.

**4.3 A Physical test**

A large mechanical oscillation occurred immediately after manually opening the upper bucket hole. After the mechanical oscillation had ceased, the maximum displayed value was of 4 grams. After the upper bucket was completely empty, the weight on the digital display varied between 2 and 3 grams. Taking the measurement drift into account, a weight variation of about +2 grams can be considered. The largest expected variation in the measured weight was calculated as about +4 grams by numerical simulation, a force with a time course similar to that in Figure 2B. Hence, the test agreed with the expected behaviour modelled by the simulation of Galileo's setup. The system was almost in equilibrium during the essay, in agreement with the report made by Galileo.

**5. Discussion**

The synthesis of the results is: the steady-state idealisation of the two-bucket experiment has the force of percussion equal to the sum of the weight of the water travelling between the two buckets and the thrust force seen by the upper bucket. The numerical simulation of the real Galilean experiment shows that the above equality does not hold exactly, although the deviation was rather small and difficult to observe with the setup he described. The physical experiment performed with a load cell instead of a two-arm balance supports the assumption that coarse measurements, such as with Galileo's two-armed balance scale, will not be able to detect the small-sized deviations in a setup weighing three or four orders of magnitude more.

The Sixth day of the Discorsi is considered an unfinished work. It was written in Arcetri in 1638 [3], during the last years of Galileo's life. The two-bucket experiment is deemed to have been performed before 1610 in Padova [3]. Accordingly to Stillman Drake [2], some notes on the water jet calculations survived. However, Palmerino [4] argues about the authenticity of such notes and does not discard forgery. In our attempt to perform a bibliographical survey of references to the two-bucket experiment only two works were identified [4,5]. The work from Hatleback [5] also present the reproduction of the experiment, however in a different scale; the natural oscillations in the two-arms balance may not give an accurate observation about the possible equilibrium.

The physical problem in the core of the Sixth Day was how to estimate the percussion, i.e., the force exerted by a fluid or a solid mass which changes speed with a complete transfer of linear momentum. The Sixth Day started with the Two-Bucket Experiment and follows with the percussion of pile drivers on stout poles inserted into the ground. Both examples are not mere toy problems, as real applications to everyday engineering exist.

A practical question emerges: what was the magnitude of the percussion, according to Galileo? He considered that the percussion was equal to the weight of the free water in the air. This conjecture was inaccurate, and the percussion could hence become highly underestimated. Here, the weight of the water in the air was formally calculated, but Galileo was unable to perform algebraic integrations. He gave a number, 10 or 12 pounds [1] (10 Tuscan pounds, corresponds to 3.39 kg [3]). However, the present calculations, considering the model of figure 1, resulted in 4.8 N or 0.49 kg, very far from his estimate. The orifice diameter of one egg remains unknown; nevertheless, an ostrich egg would be required to reach the weight reported by Galileo, and his bucket would drain very quickly.

Still in the Sixth Day, but not directly related to the two-bucket experiment, Galileo wrote: "What the future speed will then be is manifest from the things demonstrated and seen in the [discussions of the] past days. That is, the future speed will be such that, in another time equal to that of the [initial free] descent, double the space of [free] fall would be passed" [2]. This sentence gives a simple way to calculate the volume (and weight) of the water falling in the air: given the surface area of the water jet at its impact point in the lower bucket, and considering a constant fluid speed thereupon, the conoid becomes a cylinder, the latter with twice the length of the former. Mossotti [6] pointed out that Galileo knew this theorem and Mossotti employed it to calculate the percussion [6]. However, differently from Galileo, Mossotti's conoid included, in addition to the water jet, the portion of moving water that did not exert a force on the upper bucket. This additional weight corresponds to the thrust calculated in the present study. The same concept was described before by Newton in Principia Mathematica [7]. However, the straightforward estimate of the percussion made by Mossotti is valid only under the assumption that the water velocity at the surface of the upper bucket is zero.

The present development has some limitations in that effects such as friction, viscous flow, inertia and air resistance were not considered. In practice, the flowrate through the hole in the upper bucket would decrease as a result of these effects, the same holding for the thrust (or non-gravitating water). The water jet would reduce its final speed, changing the magnitude of the percussion. The extent of these effects is more difficult to estimate, but it is possible to expect forces close to equilibrium.

The fact that Galileo did not account for the thrust (or the non-gravitating water inside the upper bucket) is unexpected. He was an accurate observer and deserves the statement that he "was no doubt a master of experiment but also a genius of the cognitive expansion of experiment" [8]. Considering the setup of the experiment, the thrust effect is not clearly observed as it would be with sideways holes instead. It is possible to devise an imaginary experiment: consider two lateral, identical holes, perfectly opposite to one another in the upper bucket, instead of a single hole, and two lower buckets, each one collecting the water from one of the holes. The thrusts would cancel each other, and the sum of the percussions in the lower buckets would be equal to the weight of the water in the air as Galileo stated. However, since the water would fall in a parabolic profile, only the change in vertical linear momenta would be associated with forces applied to the structure, the horizontal projections cancelling each other. Here, a perfect symmetry is essential, making the Galilean experiment more suitable, provided, for correctness, the thrust effect is not forsaken.

In conclusion, the observations reported by Galileo from the two-bucket experiment hint at the performance of a real experiment. His clever reasoning, which includes an instantaneous transformation of the travelling water into ice, led him to conclude that the balance arms would equilibrate. Nevertheless, by stopping the water jet, two equal and contrary forces -- the thrust and part of the percussion -- would disappear. Anyway, although in consequence the percussion resulted underestimated, the disequilibrium of the two-arm balance is small enough to pass unnoticed in an experiment such as that described in Giornata Sesta of "Two New Sciences".

**Acknoledgments**

The authors thank Beatriz Lacorte Giannella for helping with the physical experiment. To the Brazilian Agencies: CAPES, CNPq and FAPERJ.